\documentclass[journal]{IEEEtran}
\usepackage{amsmath}
\usepackage{amsfonts}
\usepackage{amssymb}
\usepackage{algorithm}
\usepackage{algorithmic}
\usepackage{graphicx}
\usepackage{verbatim}

\newtheorem{theorem}{Theorem}
\newtheorem{lemma}{Lemma}

\newtheorem{example}{Example}

\newcommand{\beq}{\begin{equation}}
\newcommand{\eeq}{\end{equation}}
\newcommand{\beqnn}{\begin{equation*}}
\newcommand{\eeqnn}{\end{equation*}}
\newcommand{\beqy}{\begin{eqnarray}}
\newcommand{\eeqy}{\end{eqnarray}}
\newcommand{\beqynn}{\begin{eqnarray*}}
\newcommand{\eeqynn}{\end{eqnarray*}}
\newcommand{\bit}{\begin{itemize}}
\newcommand{\eit}{\end{itemize}}
\newcommand{\ben}{\begin{enumerate}}
\newcommand{\een}{\end{enumerate}}
\newcommand{\bex}{\begin{example}}
\newcommand{\eex}{\end{example}}

\newcommand{\balg}[1]{\begin{algorithm} \caption{#1}}
\newcommand{\ealg}{\end{algorithm}}

\newcommand{\balgc}{\begin{algorithmic}[1]}
\newcommand{\ealgc}{\end{algorithmic}}

\newcommand{\bary}{\begin{array}}
\newcommand{\eary}{\end{array}}
\newcommand{\bmx}{\begin{bmatrix}}
\newcommand{\emx}{\end{bmatrix}}
\newcommand{\bsmx}{\left[\begin{smallmatrix}}
\newcommand{\esmx}{\end{smallmatrix}\right]}
\newcommand{\bmxc}[1]{\left[\begin{array}{@{}#1@{}}}
\newcommand{\emxc}{\end{array}\right]}
\newcommand{\bcn}{\begin{center}}
\newcommand{\ecn}{\end{center}}

\newcommand{\Rbb}{{\mathbb{R}}}
\newcommand{\Zbb}{{\mathbb{Z}}}

\newcommand{\Rnbn}{\Rbb^{n \times n}}

\newcommand{\Rmbm}{\Rbb^{m \times m}}

\newcommand{\Zn}{\Zbb^{n}}

\newcommand{\sILS}{{\scriptscriptstyle ILS}}

\newcommand{\sB}{{\scriptscriptstyle B}}

\newcommand{\mbbZ}{{\mathbb{Z}}}

\newcommand{\A}{\boldsymbol{A}}

\newcommand{\D}{\boldsymbol{D}}

\newcommand{\G}{\boldsymbol{G}}

\newcommand{\I}{\boldsymbol{I}}

\renewcommand{\P}{\boldsymbol{P}}
\newcommand{\Q}{\boldsymbol{Q}}
\newcommand{\R}{\boldsymbol{R}}

\newcommand{\U}{\boldsymbol{U}}
\newcommand{\V}{\boldsymbol{V}}

\newcommand{\Z}{\boldsymbol{Z}}

\newcommand{\e}{\boldsymbol{e}}

\renewcommand{\u}{\boldsymbol{u}}
\renewcommand{\v}{\boldsymbol{v}}
\newcommand{\w}{\boldsymbol{w}}
\newcommand{\x}{{\boldsymbol{x}}}
\newcommand{\y}{{\boldsymbol{y}}}
\newcommand{\z}{\boldsymbol{z}}
\newcommand{\0}{{\boldsymbol{0}}}

\newcommand{\br}{{\bar{r}}}
\newcommand{\bby}{{\bar{\y}}}

\newcommand{\bbQ}{{\bar{\Q}}}
\newcommand{\bbR}{{\bar{\R}}}

\newcommand{\tr}{{\tilde{r}}}

\newcommand{\ty}{{\tilde{y}}}

\newcommand{\tby}{{\tilde{\y}}}

\newcommand{\hr}{{\hat{r}}}
\newcommand{\hx}{{\hat{x}}}

\newcommand{\hbG}{{\hat{\G}}}
\newcommand{\hbR}{{\hat{\R}}}

\newcommand{\hbw}{{\hat{\w}}}
\newcommand{\hbx}{{\hat{\x}}}
\newcommand{\hby}{{\hat{\y}}}
\newcommand{\hbz}{{\hat{\z}}}

\newcommand{\sign}{\mathrm{sign}}

\providecommand{\abs}[1]{| #1 |}

\usepackage{slashbox}

\begin{document}

\title{Effects of the LLL reduction on the success probability of the Babai point and
on the complexity of sphere decoding}

\author{Xiao-Wen~Chang,
        Jinming~Wen,
        and~Xiaohu~Xie% <-this % stops a space
\thanks{X.-W. Chang is with The School of Computer Science, McGill University,
Montreal, QC H3A 2A7, Canada (e-mail: chang@cs.mcgill.ca).}
\thanks{Jinming~Wen is with  The Department of Mathematics and Statistics,
McGill University, Montreal, QC H3A 0B9, Canada (e-mail: jinming.wen@mail.mcgill.ca).}
\thanks{Xiaohu~Xie is with The School of Computer Science, McGill University,
Montreal, QC H3A 2A7, Canada (e-mail: xiaohu.xie@mail.mcgill.ca).}% <-this % stops a space

\thanks{Copyright (c) 2012 IEEE. Personal use of this material is permitted.  However, permission to use this material for any other purposes must be obtained from the IEEE by sending a request to pubs-permissions@ieee.org.}}

%\markboth{}%
%{Shell \MakeLowercase{\textit{et al.}}: Bare Demo of IEEEtran.cls for Journals}


\maketitle

\begin{abstract}
%\boldmath
A common method to estimate an unknown integer parameter vector in a linear model is to solve an integer least squares (ILS) problem.
A typical approach to solving an ILS problem is sphere decoding.
To make a sphere decoder faster, the well-known LLL reduction is often used as preprocessing.
The Babai point produced by the Babai nearest plane algorithm is a suboptimal solution of the ILS problem.
First we prove that the success probability of the Babai point as a lower bound on the success probability of the ILS estimator
is sharper than the lower bound given by Hassibi and Boyd \cite{HasB98}.
Then we show  rigorously that applying the LLL reduction algorithm will increase  the success probability of the Babai point
and give some theoretical and numerical test results.
We give examples to show that unlike LLL's column permutation strategy, two often used  column permutation strategies
SQRD and V-BLAST may decrease  the success probability of the Babai point.
Finally we show rigorously that applying the LLL reduction algorithm will also reduce the  computational complexity of sphere decoders,
which is measured approximately by the number of nodes in the search tree in the literature.
\end{abstract}


% \fi
\IEEEpeerreviewmaketitle

\section{Introduction}
\IEEEPARstart{C}{onsider} the following linear model:
\begin{equation}
\label{e:model}
\y=\A\hbx+\v,
\end{equation}
where $\y\in \mathbb{R}^m$ is an observation vector,   $\A\in\mathbb{R}^{m\times n}$ is a deterministic model
matrix with full column rank,
$\hbx\in\mathbb{Z}^n$ is an unknown integer parameter vector, and
$\v\in \mathbb{R}^m$ is a noise vector following the Gaussian distribution
$\mathcal{N}(\boldsymbol{0},\sigma^2 \I)$ with $\sigma$ being known.
A common method to estimate $\hbx$ in \eqref{e:model} is to solve the following integer least squares (ILS) problem:
\beq
\label{e:ILS}
\min_{\x\in{Z}^n}\|\y-\A\x\|_2^2,
\eeq
whose solution $\x^\sILS$ is the maximum-likelihood estimator of $\hbx$.
The ILS problem is also  referred to as the closest point problem in the literature as it is equivalent to find a point in the lattice
$\{\A\x: \x\in \Zn\}$  which is closest to $\y$.

A typical approach to solving \eqref{e:ILS} is the discrete search approach,  referred to as sphere decoding in communications,
such as the Schnorr-Euchner algorithm \cite{SchnE94} or its variants, see e.g. \cite{AEVZ02,DameGC03}.
To make the search faster,  a lattice reduction is performed to transform the given problem to an equivalent problem.
A widely used  reduction is the LLL reduction  proposed by Lenstra, Lenstra and Lov\'{a}sz in \cite{LLL82}.

It has been shown that the ILS problem is NP-hard \cite{Boas81,Micci01}.
Solving \eqref{e:ILS} may become time-prohibitive when $\A$ is ill conditioned,  the noise is large,  or the dimension of the problem
is large \cite{JalO05}.
So for some applications, an approximate solution, which can be produced quickly, is  computed instead.
One often used approximate solution is the Babai  point, produced by Babai's nearest plane algorithm
\cite{Bab86}. This  approximate solution is also the first integer point found by  the Schnorr-Euchner algorithm.
In communications, a method for finding this approximate solution is referred to as a successive interference cancelation decoder.

In order to verify whether an estimator is good enough for a practical use,
one needs to find the probability of the estimator being equal to the true integer parameter vector, which is referred to as success probability \cite{HasB98}.
The probability of wrong estimation is referred to as error probability, see, e.g., \cite{JaldB09}.

If  the Babai  point is used as an estimator of the integer parameter vector $\hbx$ in \eqref{e:model}, certainly
it is important to find its success probability, which can easily be computed.
Even if one intends to compute the ILS estimator, it is still important to find  the success probability
of  the Babai  point.
It is very difficult to compute the success probability of the ILS estimator, so  lower and upper bounds have been
considered to approximate it, see, e.g., \cite{HasB98, Xu06}.
In \cite{Teun99} it was shown that the success probability of the ILS estimator
is the largest among all ``admissible'' estimators, including the Babai  point, which is referred to
as a bootstrapping estimator in  \cite{Teun99}.
The  success probability of the Babai point is often used as an approximation to  the success probability of the ILS estimator.
In general, the higher the success probability of the Babai point, the lower the complexity of finding the ILS estimator by the
discrete search approach.
In practice, if the success probability of the Babai point is high, say close to 1, then one does not need to spend extra computational
time to find the ILS estimator.

Numerical experiments have shown that after the LLL reduction, the  success probability of the Babai point increases \cite{GanM08}.
But whether the  LLL reduction can always improve the success probability of the Babai point  is still unknown.
In this paper, we will prove that  the success probability of the Babai point will become higher after the LLL reduction algorithm is used.
It is well-known that the LLL reduction can make sphere decoders faster.
But to our knowledge there is still no rigorous justification.
We will show that the LLL reduction can always decrease the computational complexity of sphere decoders,
an approximation to the number of nodes in the search tree given in the literature.

The rest of the paper is organized as follows.
In section \ref{s:lll}, we introduce the LLL reduction to reduce the ILS probelm \eqref{e:ILS}.
In section \ref{s:babai}, we introduce the Babai  point and a formula to compute the success probability of the Babai point,
and we show that the success probability of the Babai point is a sharper lower bound on  the success probability of ILS estimator
compared with the lower bound given in \cite{HasB98}.
In section \ref{s:prob}, we rigorously prove that the LLL reduction algorithm improves the success probability of the Babai point.
 In section \ref{s:comp}, we rigorously show that the LLL reduction algorithm reduces the computational complexity of sphere decoders. Finally  we summarize this paper in section \ref{s:sum}.

In this paper, $\e_k$ denotes the $k$-th column of the identity matrix $\I$.
For $\x\in \mathbb{R}^n$, we use $\lfloor \x\rceil$ to denote its nearest integer vector, i.e.,
each entry of $\x$ is rounded to its nearest integer (if there is a tie, the one with smaller magnitude is chosen).
For a vector $\x$, $\x_{i:j}$ denotes the subvector of $\x$ formed by entries $i, i+1, \ldots,j$.
For a matrix $\A$, $\A_{i:j,i:j}$ denotes the submatrix of $\A$ formed by rows and columns $i, i+1, \ldots,j$.
The success probabilities of the Babai point and the ILS estimator are denoted by $P_{\sB}$ and $P_{\sILS}$, respectively.

\section{LLL Reduction and transformation of the ILS Problem}\label{s:lll}

Assume that $\A$ in the linear model \eqref{e:model} has the QR factorization
$$
\A=[\Q_1, \Q_2]\bmx\R \\ \0 \emx,
$$
where $[\underset{n}{\Q_1}, \underset{m-n}{\Q_2}]\in \Rmbm$ is orthonormal and $\R\in \Rnbn$ is upper triangular.
Without loss of generality, we assume the diagonal entries of $\R$ are positive throughout the paper.
Define $\tby=\Q_1^T\y$.
From \eqref{e:model},  we have $\tby=\R\hbx+\Q_1^T \v$.
Because $\v\sim \mathcal{N}(\boldsymbol{0},\sigma^2 \I)$,
it follows that $\tby \sim \mathcal{N}(\R\hbx, \sigma^2 \I)$.

With the QR factorization of $\A$,   the ILS problem \eqref{e:ILS} can be transformed to
\beq
\label{e:ILSR}
\min_{\x\in{Z}^n}\|\tby-\R\x\|_2^2.
\eeq
One can then apply a sphere decoder such as the Schnorr-Euchner search algorithm \cite{SchnE94}
to find the solution of \eqref{e:ILSR}.

The efficiency of the search process depends on $\R$.
For efficiency, one typically uses the LLL reduction instead of the QR factorization.
After the QR factorization of $\A$, the LLL reduction \cite{LLL82} reduces the matrix $\R$ in \eqref{e:ILSR} to $\bbR$:
\beq
\label{e:QRZ}
\bbQ^T \R \Z = \bbR,
\eeq
where $\bbQ  \in \mathbb{R}^{n\times n}$ is orthonormal,
$\Z\in   \mathbb{Z}^{n\times n}$ is a  unimodular matrix (i.e., $\det(\Z)=\pm1$),
and  $\bbR\in \mathbb{R}^{n\times n}$ is upper triangular with positive diagonal entries and satisfies the following conditions:
\begin{align}
&|\br_{ik}|\leq\frac{1}{2}\br_{ii}, \quad i=1, 2, \ldots, k-1 \label{e:criteria1} \\
&\delta \br_{k-1,k-1}^2 \leq   \br_{k-1,k}^2+ \br_{kk}^2,\quad k=2, 3, \ldots, n, \label{e:criteria2}
\end{align}
where $\delta$ is a constant satisfying $1/4 < \delta \leq 1$.
The matrix $\R$ is said to be $\delta$-LLL reduced or simply LLL reduced.
%In practice $\delta$ is often set to 1 in solving ILS problems.
Equations \eqref{e:criteria1} and  \eqref{e:criteria2} are referred to as the size-reduced condition and
the Lov\'{a}sz condition, respectively.

The original LLL algorithm given in \cite{LLL82} can be described in the matrix language.
Two types of basic unimodular matrices are  implicitly used to update $\R$
so that it satisfies the two conditions.
One is  the integer Gauss transformations (IGT) matrices and the other is permutation matrices,
see below.

To meet the first condition in \eqref{e:criteria1}, we can apply an IGT, which has  the following form:
\beqnn
\Z_{ik}=\I-\zeta \e_i\e_k^T.
\eeqnn
Applying $\Z_{ik}\ (i < k)$ to $\R$ from the right gives
\beqnn
\bbR= \R\Z_{ik} = \R-\zeta \R\e_i\e_k^T.
\eeqnn
Thus $\bbR$ is the same as $\R$, except that $\br_{jk} = r_{jk} - \zeta r_{ji}$ for $j = 1,   \ldots, i$.
By setting $\zeta= \lfloor r_{ik}/r_{ii} \rceil $, we ensure $|\br_{ik}|\le\br_{ii}/2$.

To meet the second condition in  \eqref{e:criteria2} permutations are needed in the reduction process.
Suppose that  $\delta\, r_{k-1,k-1}^2 > r^2_{k-1,k}+r^2_{k,k}$ for some $k$.
Then we interchange columns $k-1$ and $k$ of $\R$.
After the permutation  the upper triangular structure of $\R$ is no longer maintained.
But we can bring $\R$ back to an upper triangular matrix by using the Gram-Schmidt orthogonalization technique
(see \cite{LLL82}) or by a Givens rotation:
\beq
\label{e:Permu}
\bbR=\G_{k-1,k}^T\R\P_{k-1,k},
\eeq
where $\G_{k-1,k}$ is an orthonormal matrix and $\P_{k-1,k}$ is a permutation matrix,  and
\begin{align}
\br_{k-1,k-1}^2&= r^2_{k-1,k}+r^2_{k,k}, \nonumber\\
\br^2_{k-1,k}+\br^2_{k,k}&= r_{k-1,k-1}^2. \label{e:PerR}
\end{align}
Note that the above operation  guarantees $\delta\, \br_{k-1,k-1}^2 <  \br^2_{k-1,k}+\br^2_{k,k}$ since $\delta \leq 1$.
The LLL reduction algorithm is described in Algorithm \ref{a:LLL}, where
the final reduced upper triangular matrix is still denoted by $\R$.

\begin{algorithm}[h!]
\caption{LLL reduction}   \label{a:LLL}
\begin{algorithmic}[1]
  \STATE compute the QR factorization: $\A=\Q\bmx \R \\ \0 \emx$;
  \STATE set $\Z=\I_n$, $k=2$;
  \WHILE{$k\le n$}
   \STATE apply IGT $\Z_{k-1,k}$ to reduce $r_{k-1,k}$: \\
    \ \ \ \ $\R=\R\Z_{k-1,k}$;
   \STATE update $\Z$: $\Z=\Z\Z_{k-1,k}$;
   \IF{$\delta\,  r_{k-1,k-1}^2>  r^2_{k-1,k}+r^2_{kk}$}
    \STATE permute and triangularize $\R$: \\
     \ \ \ \ $\R\!=\!\G_{k-1,k}^T\R\P_{k-1,k}$; \label{l:pt}
    \STATE update $\Z$: $\Z=\Z\P_{k-1,k}$;
    \STATE $k=k-1$, when $k>2$;
   \ELSE
    \FOR{$i=k-2,\dots,1$}
      \STATE apply IGT $\Z_{ik}$ to reduce $r_{ik}$: $\R=\R\Z_{ik}$;
      \STATE update $\Z$: $\Z=\Z\Z_{i,k}$;
    \ENDFOR
    \STATE $k=k+1$;
   \ENDIF
  \ENDWHILE
%  \STATE set $\bbR=\R$;
\end{algorithmic}
\end{algorithm}

After the LLL reduction \eqref{e:QRZ}, the ILS problem \eqref{e:ILSR} is then transformed to:
\beq
\label{e:reduced}
\min_{\z\in{Z}^n}\|\bby-\bbR\z\|_2^2,
\eeq
where $\bby=\bbQ^T\tby$ and $\z=\Z^{-1}\x$.

The LLL reduction is a powerful preprocessing tool that allows to reduce the complexity of
search process for finding the ILS solution, see, e.g., \cite{HasB98,AEVZ02}.

\section{Success Probability of the Babai  point and a lower bound}\label{s:babai}

The Babai  (integer) point $\x^\sB\in \Zbb^n$ found by the Babai nearest plane algorithm  \cite{Bab86}
is defined as follows:
\beq \label{e:ck}
\begin{split}
& c_n=\ty_n/r_{nn}, \quad x_n^\sB=\lfloor c_n\rceil, \\
& c_{i}=(\ty_{i}-\sum_{j=i+1}^nr_{ij}x_j^\sB)/r_{ii}, \quad  x_i^\sB=\lfloor c_i\rceil,
\end{split}
\eeq
for $i= n-1, \ldots, 1.$
Note that the entries of $\x^\sB$ are determined from the last to the first.
The Babai  point $\x^\sB$  is actually the first integer point found by the Schnorr-Euchner search algorithm \cite{SchnE94}
for solving \eqref{e:ILSR}.

In the following we give a formula for the success probability of the Babai  point.
The formula is equivalent to the one given by Teunissen in \cite{Teun981}, which
considers  a variant form of the ILS problem  \eqref{e:ILS}.
But our proof is easier to follow than that given in \cite{Teun981}.

\begin{theorem} \label{t:prob}
Suppose $\tby \sim {\cal N}(\R\hbx, \sigma^2\I)$ in the ILS problem \eqref{e:ILSR}.
Let $P_\sB$ denotes the success probability  of the Babai  point  $\x^\sB$ given in  \eqref{e:ck},
i.e., $P_\sB=\Pr(\x^\sB=\hbx)$.
Then
\beq
P_\sB  =\prod_{i=1}^n\phi(r_{ii}), \;
\phi(\zeta)=\sqrt{\frac{2}{\pi}}\int_0^{\zeta/(2\sigma)}\exp(-\frac{1}{2}t^2)dt.  \label{e:pb}
\eeq
\end{theorem}

{\em Proof.} By the chain rule of conditional probabilities:
\begin{align}
P_\sB & = \Pr(\x^{\sB}=\hbx)  =P\big(\bigcap_{i=1}^n(x_i^\sB=\hx_i)\big) =\Pr(x_n^\sB=\hx_{n}) \nonumber \\
& \times \prod_{i=1}^{n-1}\Pr(x_i^\sB=\hx_i|x_{i+1}^\sB=\hx_{i+1}, \cdots, x_n^\sB=\hx_{n}).
 \label{e:chain}
\end{align}

Since $\tby\sim \mathcal{N}(\R\hbx,\sigma^2 \I)$, we have
\begin{align*}
& \ty_n \sim\mathcal{N}(r_{nn}\hx_n, \sigma^2), \\
& \ty_i \sim\mathcal{N}(r_{ii}\hx_{i}+\sum_{j=i+1}^n r_{ij}\hx_j, \sigma^2), \quad i=n-1,\ldots, 1.
\end{align*}
Thus, from \eqref{e:ck} we have
$$
c_n   \sim\mathcal{N}(\hx_{n},  \sigma^2/r_{nn}^2 ),
$$
and if $ x_{i+1}^\sB=\hx_{i+1}, \cdots, x_n^\sB=\hx_{n}$,
$$
c_i  \sim\mathcal{N}(\hx_{i},  \sigma^2/r_{ii}^2).
$$
Then it follows that
\begin{align*}
\Pr(x_n^\sB =\hx_{n})
&= \Pr(|c_n-\hx_{n}| \leq 1/2) \\
&=\frac{1}{\sqrt{2\pi}\frac{\sigma}{r_{nn}}}
\int_{-0.5}^{0.5}\exp(-\frac{t^2}{2(\frac{\sigma}{r_{nn}})^2})dt\nonumber\\
& =\frac{2}{\sqrt{2\pi}}\int_0^{r_{nn}/(2\sigma)}\exp(-\frac{1}{2}t^2)dt=\phi(r_{nn}).
\end{align*}
Similarly, we can obtain
\begin{eqnarray*}
&\Pr(x_i^\sB=\hx_{i}|x_{i+1}^\sB=\hx_{i+1}, \cdots,
x_{n}^\sB=\hx_{n})=\phi(r_{ii}).
\end{eqnarray*}
Then from \eqref{e:chain} we can conclude that \eqref{e:pb} holds. \ \ $\Box$
\medskip

Since $P_\sB$ in \eqref{e:pb} depends on $\R$, sometimes we also write $P_\sB$ as $P_\sB(\R)$.

The success probability $P_{\sILS}$ of the ILS estimator depends on its Voronoi cell \cite{HasB98}
and it is difficult to compute it because the shape of Voronoi cell is complicated.
In \cite{HasB98} a lower bound $F(d_{\min}^2/(4\sigma^2),n)$
is proposed to approximate it,
where $d_{\min}$  is the length of the shortest lattice vector, i.e., $d_{\min}=\min_{\0 \neq \x \in \Zn}\|\R\x\|_2$,
and $F$ is the cumulative distribution function of chi-square distribution.
However, no polynomial-time algorithm has been found to compute $d_{\min}$.
To overcome this problem,  \cite{HasB98} proposed a more  practical lower bound $F(r_{\min}^2/(4\sigma^2),n)$,
where $r_{\min}\equiv \min_i r_{ii}$.
Note that $P_\sB$ is also a lower bound on $P_{\sILS}$ (see \cite{Teun99}).
The following result shows that $P_\sB$ is sharper than $F(r_{\min}^2/(4\sigma^2),n)$.

\begin{theorem}
$
F\Big(\frac{r_{\min}^2}{4\sigma^2},n\Big)\leq P_\sB.
$
\end{theorem}

{\em Proof.} Let $\u \sim {\cal N}(\0, \I_n)$. Thus $u_1, u_2, \ldots, u_n$ are i.i.d. and $\sum_{i=1}^n u_i^2$ follows the chi-squared distribution with degree $n$.
Let events $E=\{\sum_{i=1}^n u_i^2\leq r_{\min}^2/(4\sigma^2)\}$
and $E_i=\{u_i^2\leq r_{ii}^2/(4\sigma^2)\}$ for $i=1,2,\ldots, n$.
Since $r_{\min}\leq r_{ii}$,   $E\subseteq \bigcap_{i=1}^n E_i$.
Thus,
\begin{align*}
F\Big(\frac{r_{\min}^2}{4\sigma^2},n\Big)
& = \Pr(E) \leq \Pr(\bigcap_{i=1}^n E_i) =\prod_{i=1}^n \Pr(E_i) \\
& = \prod_{i=1}^n \frac{1}{\sqrt{2\pi}}\int_{-r_{ii}/(2\sigma)}^{r_{ii}/(2\sigma)}\exp\big(-\frac{1}{2}t^2\big)dt \\
& =  \prod_{i=1}^n \phi(r_{ii}) = P_\sB. \ \ \ \  \hfill \Box
\end{align*}
\smallskip

In the following, we give an  example to show that $F(r_{\min}^2/(4\sigma^2),n)$ can be much smaller than $P_\sB$.
\bex \label{ex:lbd}
Let $\R=\bmx 0.001   & 0\\ 0 &  10\emx$ and $\sigma=0.5$.
By simple calculations, we obtain
$F(r_{\min}^2/(4\sigma^2),n)/P_\sB=1/1596$.
Although this is a contrived example, where the signal-to-noise ratio is  small,
it shows that  $P_\sB$ can be much sharper than  $F(r_{\min}^2/(4\sigma^2),n)$
as a lower bound on $P_\sILS$.
\eex

%%%%%%%%%%%%%%%%%%%%%%%%%%%%%%%%%%%%%%%%%%%%%%%%%%%%%%%%%%%%%%%
\section{Enhancement of $P_\sB$ by the LLL reduction}\label{s:prob}

In this section we rigorously prove that column permutations and size reductions in the LLL reduction process given in Algorithm \ref{a:LLL}
enhance  (not strictly) the success probability $\P_\sB$ of the Babai  point.
We give simulations to show that unlike LLL's column permutation strategy, two often used  column permutation strategies
SQRD \cite{WBRKK} and V-BLAST \cite{FGVW} may decrease  the success probability of the Babai point.
We will also discuss how the parameter $\delta$ affects the enhancement and give some upper bounds on  $\P_\sB$
after the LLL reduction.

%%%%%%%%%%%%%%%%%%%%%%%%%%%%%%%%%%%%%
\subsection{Effects of the LLL reduction on $P_\sB$ } \label{s:lllprob}
Suppose that we have the QRZ factorization \eqref{e:QRZ}, where $\bbQ$ is orthonormal, $\Z$ is unimodular
and $\bbR$ is upper triangular with positive diagonal entries (we do not assume that $\bbR$ is LLL reduced
unless we state otherwise).
Then with $\bby=\bbQ^T\tby$ and $\z=\Z^{-1}\x$ the ILS problem \eqref{e:ILSR} can be transformed to \eqref{e:reduced}.
For \eqref{e:reduced} we can also define its corresponding Babai  point $\z^\sB$.
This Babai  point  can be used as an estimator of $\hbz \equiv \Z^{-1}\hbx$, or equivalently $\Z\z^\sB$
can be used an estimator of $\hbx$.
In  \eqref{e:ILSR} $\tby\sim {\cal N}(\R\hbx, \sigma^2\I)$.  It is easy to verify that
in \eqref{e:reduced} $\bby \sim {\cal N}(\bbR\hbz, \sigma^2\I)$.
In the following we look at how the success probability of the Babai  point changes after some specific  transformation is used to $\R$.

The following result shows that if the Lov\'{a}sz condition \eqref{e:criteria2} is not satisfied,
after a column permutation and triangularization, the success probability of the Babai  point increases.

\begin{lemma}\label{l:probper}

Suppose that $\delta\, r_{k-1,k-1}^2 > r^2_{k-1,k}+r^2_{kk}$ for some $k$ for the $\R$ matrix in the ILS problem \eqref{e:ILSR}.
After the permutation of columns $k-1$ and $k$ and triangularization, $\R$ becomes $\bbR$,
i.e.,  $\bbR=\G_{k-1,k}^T\R \P_{k-1,k}$ {\rm (see \eqref{e:Permu})}.
With $\bby=\G_{k-1,k}^T\tby$ and $\z=\P_{k-1,k}^{-1}\x$,
 \eqref{e:ILSR} can be transformed to \eqref{e:reduced}.
Denote $\hbz \equiv \P_{k-1,k}^{-1}\hbx$.
Then the Babai point $\z^\sB$ has a success probability greater than or equal to  the Babai  point $\x^\sB$, i.e.,
\beq
\Pr(\x^\sB=\hbx) \leq \Pr(\z^\sB=  \hbz),
\label{e:probper}
\eeq
where the equality holds if and only if $r_{k-1,k}=0$.
\end{lemma}

{\em Proof.} By Theorem \ref{t:prob},  what we need to show is the following inequality:
\begin{equation}
\prod_{i=1}^n\phi(r_{ii})\leq \prod_{i=1}^n\phi(\br_{ii}).
\label{e:pineq}
\end{equation}
Since $\br_{ii}=r_{ii}$ for $i\neq k-1,k$, we  only need to show
$$
\phi(r_{k-1,k-1})\phi(r_{kk}) \leq \phi(\br_{k-1,k-1})\phi(\br_{kk}),
$$
which  is equivalent to
\beq \label{e:ineq}
\begin{split}
 & \int_0^{\frac{r_{k-1,k-1}}{2\sigma}} \exp(-\frac{1}{2}t^2)dt
 \int_0^{\frac{r_{kk}}{2\sigma}} \exp(-\frac{1}{2}t^2)dt \\
\leq &  \int_0^{\frac{\br_{k-1,k-1}}{2\sigma}} \exp(-\frac{1}{2}t^2)dt
 \int_0^{\frac{\br_{kk}}{2\sigma}} \exp(-\frac{1}{2}t^2)dt.
\end{split}
\eeq

Since  $\G_{k-1,k}$ is orthonormal and $\P_{k-1,k}$ is a permutation matrix,
the absolute value of the determinant of the submatrix $\R_{k-1:k,k-1:k}$ is unchanged, i.e.,
we have
\beq
r_{k-1,k-1}r_{kk}=\br_{k-1,k-1}\br_{kk}.
\label{e:det}
\eeq

Let
\begin{align}
& a=\frac{r_{k-1,k-1}}{2\sigma}\frac{r_{kk}}{2\sigma}=\frac{\br_{k-1,k-1}}{2\sigma}\frac{\br_{kk}}{2\sigma},
  \label{e:prod} \\
& f(\zeta)=\ln\int_0^{\zeta}\exp(-\frac{1}{2}t^2)dt+\ln\int_0^{{a}/{\zeta}}\exp(-\frac{1}{2}t^2)dt. \label{e:fun}
\end{align}
Note that $f(\zeta)=f(a/\zeta)=f(\max\{\zeta, a/\zeta\})$. Then  \eqref{e:ineq} is equivalent to
\beq
f\Big(\frac{\max\{r_{k-1,k-1}, r_{kk}\}}{2\sigma}\Big) \leq f\Big(\frac{\max\{\br_{k-1,k-1},\br_{kk}\}}{2\sigma}\Big).
\label{e:inqu1}
\eeq

Obviously, if $r_{k-1,k}=0$, then the equality in \eqref{e:inqu1} holds since in this case
$$\frac{\max\{r_{k-1,k-1}, r_{kk}\}}{2\sigma}=\frac{\max\{\br_{k-1,k-1},\br_{kk}\}}{2\sigma}.$$
So we only need to
show if $r_{k-1,k}\neq0$, then the strict inequality in \eqref{e:inqu1} holds. In the following,
we assume $r_{k-1,k}\neq0$.

From $\delta r_{k-1,k-1}^2  > r_{k-1,k}^2+r_{kk}^2$  and \eqref{e:PerR} we can conclude that
$$
r_{kk}, \br_{k-1,k-1},  \br_{kk}  <  r_{k-1,k-1}.
$$
Then, with \eqref{e:prod} it follows that
\begin{align*}
&\frac{\max\{r_{k-1,k-1}, r_{kk}\}}{2\sigma}
=\frac{r_{k-1,k-1}}{2\sigma}\\
&>\frac{\max\{\br_{k-1,k-1},\br_{kk}\}}{2\sigma}\geq \sqrt{a}.
\end{align*}
Thus, to show the strict  inequality in \eqref{e:inqu1} holds, it suffices to show that when $\zeta>\sqrt{a}$, $f(\zeta)$ is a strict monotonically decreasing function
or equivalently $f'(\zeta)<0$.

From \eqref{e:fun},
\begin{align*}
f'(\zeta)
& =\frac{\exp(-\frac{1}{2}\zeta^2)}{\int_0^{\zeta}\exp(-\frac{1}{2}t^2)dt}-
\frac{\frac{a}{\zeta^2}\exp(-\frac{(a/\zeta)^2}{2})}{\int_0^{a/\zeta}\exp(-\frac{1}{2}t^2)dt} \\
& = \frac{1}{\zeta}\left (g(\zeta)-g\Big(\frac{a}{\zeta}\Big)\right),
\end{align*}
where $g(\zeta)=\frac{\zeta \exp(-\frac{1}{2}\zeta^2)}{\int_0^{\zeta}\exp(-\frac{1}{2}t^2)dt}$.
%Note that $f'(\sqrt{a})=0$.
Note that $\zeta > \sqrt{a}$,  $\zeta>a/\zeta$.
Thus, in order to show $f'(\zeta)<0$ for $\zeta>\sqrt{a}$,
we need only to show that $g(\zeta)$ is a strict monotonically decreasing function
or equivalently $g'(\zeta)<0$ when $\zeta>0$.

Simple calculations give
\begin{align*}
g'(\zeta)=& \frac{\exp(-\frac{1}{2}\zeta^2)}{(\int_0^{\zeta}\exp(-\frac{1}{2}t^2)dt)^2} \\
& \times \left[(1-\zeta^2)\int_0^{\zeta}\exp(-\frac{1}{2}t^2)dt-\zeta\exp(-\frac{1}{2}\zeta^2)\right].
\end{align*}
If $1-\zeta^2 \leq 0$ and $\zeta>0$, then obviously $g'(\zeta)<0$.
If $1-\zeta^2>0$ and $\zeta>0$,  since $\exp(-\frac{1}{2}t^2)\leq 1$,
\begin{equation*}
(1-\zeta^2)\int_0^{\zeta}\exp(-\frac{1}{2}t^2)dt\leq \zeta(1-\zeta^2) < \zeta\exp(-\frac{1}{2}\zeta^2),
\end{equation*}
where the second inequality can easily be verified. Thus again   $g'(\zeta)<0$ when  $\zeta>0$,
completing the proof.    \ \ $\Box$
\medskip

Now we make some remarks.
The above proof shows  that $f(\zeta)$ for $\zeta \geq \sqrt{a}$ reaches its maximum when $\zeta=\sqrt{a}$.
Thus if $\br_{k-1,k-1}=\br_{kk}$, or equivalently,
$$
r_{k-1,k}^2+r_{kk}^2=r_{k-1,k-1}r_{kk},
$$
$P_\sB$ will increase most.
For a more general result, see Lemma \ref{l:upp} and the remark after it.

In Lemma \ref{l:probper} there is no requirement that $r_{k-1,k}$ should be size-reduced.
The question we would like to ask here is do size reductions in the LLL reduction algorithm affect $P_\sB$?
From \eqref{e:pb} we observe that $P_\sB$ only depends on the diagonal entries of $\R$.
Thus size reductions {\em alone} will not change $P_\sB$.
However, if a size reduction can bring   changes to the diagonal entries of $\R$ after a permutation,
then it will likely affect  $P_\sB$.
Therefore, all the size reductions on the off-diagonal entries above the superdiagonal have no effect on $P_\sB$.
But the size reductions on the superdiagonal entries may  affect $P_\sB$.
There are a few different situations, which we will discuss below.

Suppose that the Lov\'{a}sz condition \eqref{e:criteria2} holds for a specific $k$.
If \eqref{e:criteria2} does not hold any more  after the size reduction on $r_{k-1,k}$,
then columns $k-1$ and $k$ of $\R$ are permuted by the LLL reduction algorithm
and according to Lemma \ref{l:probper} $P_\sB$ strictly increases or keeps unchanged
if and only if the size reduction makes $r_{k-1,k}$ zero (this occurs if  $r_{k-1,k}$ is a multiple of $r_{k-1,k-1}$
before the reduction).
If \eqref{e:criteria2} still holds after the size reduction on $r_{k-1,k}$,
then this size reduction does not affect $P_\sB$.

Suppose that the Lov\'{a}sz condition \eqref{e:criteria2} does not hold for a specific $k$.
Then by Lemma \ref{l:probper} $P_\sB$ increases after a permutation and triangularization.
If the size reduction on $r_{k-1,k}$ is performed before the permutation,
we show in the next lemma that  $P_\sB$ increases further.

\begin{lemma}\label{l:probsr}
 Suppose  that   in the ILS problem \eqref{e:ILSR} $\R$ satisfies
 $\delta\, r_{k-1,k-1}^2 > r^2_{k-1,k}+r^2_{kk}$ and $|r_{k-1,k}|>r_{k-1,k-1}/2$ for some $k$.
Let $\bbR$, $\bby$, $\z$ and $\hbz$ be defined as in Lemma \ref{l:probper}.
Suppose  a size reduction on $r_{k-1,k}$ is performed first and then after the permutation of columns $k-1$ and $k$ and
 triangularization, $\R$ becomes $\hbR$, i.e.,  $\hbR=\hbG_{k-1,k}^T\R \Z_{k-1,k}\P_{k-1,k}$.
Let $\hby=\hbG_{k-1,k}^T\tby$ and $\w=\P_{k-1,k}^{-1}\Z_{k-1,k}^{-1}\x$,
then \eqref{e:ILSR} is transformed to $\min_{\w\in \Zn}\|\hby-\hbR\w\|_2$.
Denote $\hbw= \P_{k-1,k}^{-1}\Z_{k-1,k}^{-1}\hbx$.
Then the Babai point $\w^\sB$ corresponding to the new transformed ILS problem
 has a success probability greater than or equal to   the Babai  point $\z^\sB$, i.e.,
\beq
\Pr(\z^\sB= \hbz) \leq \Pr(\w^\sB= \hbw),
\label{e:probsr}
\eeq
where the equality holds if and only if
\beq
|r_{k-1,k-1}r_{k-1,k}|=r_{k-1,k}^2+r_{kk}^2.
\label{e:condition}
\eeq
\end{lemma}

{\em Proof.} Obviously \eqref{e:probsr} is equivalent to
$$
\phi(\br_{k-1,k-1})\phi(\br_{kk})\leq \phi(\hr_{k-1,k-1})\phi(\hr_{kk}),
$$
which, by the proof of Lemma \ref{l:probper}, is also equivalent to
$$
f\Big(\frac{\max\{ \br_{k-1,k-1}, \br_{kk}\}}{2\sigma}\Big) \leq f\Big(\frac{\max\{ \hr_{k-1,k-1}, \hr_{kk}}{2\sigma}\}\Big),
$$
where $f$ is defined in \eqref{e:fun}.
Since $f(\zeta)$ has been showed to be strict monotonically decreasing when $\zeta > \sqrt{a}$,
what we need to show is that
\beq
\max\{ \br_{k-1,k-1}, \br_{kk}\} \geq \max\{ \hr_{k-1,k-1}, \hr_{kk}\},
\label{e:rkm1k}
\eeq
where the equality holds if and only if \eqref{e:condition} holds.

Since $|r_{k-1,k}|>r_{k-1,k-1}/2$,
\begin{align*}
& \br_{k-1,k-1}=\sqrt{r_{k-1,k}^2+r_{kk}^2} > \sqrt{r_{k-1,k-1}^2/4+r_{kk}^2},   \\
& \br_{kk}= \frac{r_{k-1,k-1}r_{kk}}{\sqrt{r_{k-1,k}^2+r_{kk}^2}} <  \frac{r_{k-1,k-1}r_{kk}}{\sqrt{r_{k-1,k-1}^2/4+r_{kk}^2}}.
\end{align*}
But $\sqrt{r_{k-1,k-1}^2/4+r_{kk}^2} \geq  \frac{r_{k-1,k-1}r_{kk}}{\sqrt{r_{k-1,k-1}^2/4+r_{kk}^2}}$, thus
$$
\max\{ \br_{k-1,k-1}, \br_{kk}\} =  \br_{k-1,k-1}.
$$
Suppose that after the size reduction, $r_{k-1,k}$ becomes $\tr_{k-1,k}$.
Note that
\begin{align*}
\hr_{k-1,k-1}\!=\!\sqrt{{\tr_{k-1,k}}^2+r_{kk}^2}\!<\!\sqrt{r_{k-1,k}^2+r_{kk}^2}\!=\!\br_{k-1,k-1}.
%\label{e:hbkm1}
\end{align*}
Thus, it follows from \eqref{e:rkm1k} what we need to prove is that $\hr_{kk}\leq \br_{k-1,k-1}$ or equivalently
\begin{align}\label{e:equality}
\hr_{kk}\leq \sqrt{r_{k-1,k}^2+r_{kk}^2},
\end{align}
and the equality holds if and only if \eqref{e:condition} holds.

By the conditions given in the lemma,
$$
|r_{k-1,k}|< r_{k-1,k-1} < 2 |r_{k-1,k}|.
$$
Thus
\begin{align*}
\tr_{k-1,k}
& =r_{k-1, k}-\lfloor r_{k-1,k}/r_{k-1,k-1} \rceil r_{k-1,k-1}  \\
& = r_{k-1,k}-\sign(r_{k-1,k}) r_{k-1,k-1}.
\end{align*}
Now we consider two cases $r_{k-1,k}>0$ and $r_{k-1,k}<0$ separately.
If $r_{k-1,k}>0$, then
\begin{align*}
\hr_{kk}
& = \frac{r_{k-1,k-1}r_{kk}}{\hr_{k-1,k-1}}=\frac{r_{k-1,k-1}r_{kk}}{\sqrt{{\tr_{k-1,k}}^2+r_{kk}^2}} \\
& =\frac{r_{k-1,k-1}r_{kk}}{\sqrt{(r_{k-1,k}- r_{k-1,k-1})^2+r_{kk}^2}}.
\end{align*}
Thus, to show \eqref{e:equality} it suffices to show that
\begin{align*}
\frac{r_{k-1,k-1}r_{kk}}{\sqrt{(r_{k-1,k}- r_{k-1,k-1})^2+r_{kk}^2}} \leq \sqrt{r_{k-1,k}^2+r_{kk}^2}.
\end{align*}
Simple algebraic manipulations shows that the above inequality is equivalent to
$$
(r_{k-1,k-1}r_{k-1,k}-r_{k-1,k}^2-r_{kk}^2)^2 \geq 0,
$$
which certainly holds. And obviously, the equality in \eqref{e:equality} holds  if and only if
$$
r_{k-1,k-1}r_{k-1,k}=r_{k-1,k}^2+r_{kk}^2.
$$
If $r_{k-1,k}<0$, we can similarly prove that \eqref{e:equality} holds and the
equality holds if and only if
$$-r_{k-1,k-1}r_{k-1,k}=r_{k-1,k}^2+r_{kk}^2,$$
completing the proof. \ \ $\Box$
\medskip

Here we make a remark about the equality \eqref{e:condition}.
From the proof of Lemma \ref{l:probsr} we see that if \eqref{e:condition} holds, then the equality in \eqref{e:equality} holds,
thus $\hr_{kk}=\br_{k-1,k-1}$.
But the absolute value of the determinant of the submatrix $\R_{k-1:k,k-1:k}$ is unchanged
by the size reduction,  we must have $\hr_{k-1,k-1}=\br_{kk}$.
Thus if \eqref{e:condition} holds, the effect of the size reduction on $r_{k-1,k}$
is to make $\br_{k-1,k-1}$ and $\br_{kk}$ permuted; therefore the success probability $P_\sB$
is not changed by the size reduction.
Here we give an example.

\bex
Let $\R=\bmx 5 & 4 \\ 0 & 2 \emx$. Then it is easy to verify that $\bbR=\bmx 2\sqrt{5} & 2\sqrt{5} \\ 0 & \sqrt{5} \emx$
and $\hbR=\bmx \sqrt{5} & -\sqrt{5} \\ 0 & 2\sqrt{5} \emx$.
From the diagonal entries of $\bbR$ and $\hbR$ we can conclude that the success probabilities of the two  Babai points
corresponding to $\bbR$ and $\hbR$ are equal.
\eex
\medskip

From Lemmas \ref{l:probper} and \ref{l:probsr} we immediately obtain the following results.

\begin{theorem}\label{t:problll}
Suppose that the ILS problem \eqref{e:ILSR} is transformed to the ILS problem \eqref{e:reduced},
where $\bbR$ is obtained by Algorithm \ref{a:LLL}.
Then
\beqnn
\Pr(\x^\sB=\hbx) \leq \Pr(\z^\sB=\hbz),
\eeqnn
where the equality holds if and only if no column permutation occurs during the LLL reduction process
or whenever two consecutive columns, say $k-1$ and $k$, are permuted,
$r_{k-1,k}$ is a multiple of $r_{k-1,k-1}$ (before the size reduction on $r_{k-1,k}$ is performed).
Any size reductions on the superdiagonal entries of $\R$ which are immediately followed by a column permutation
during the LLL reduction process will enhance the success probability of the Babai  point.
All other size reductions have no effect on the success probability of the Babai  point.
 \end{theorem}

\medskip

Now we make some remarks.
Note that the LLL reduction is not unique. Two different LLL reduction algorithms may produce different $\R$'s.
In Algorithm \ref{a:LLL}, when the Lov\'{a}sz condition for two consecutive columns is not satisfied,
then a column permutation takes places to ensure the Lov\'{a}sz condition to be satisfied.
If an algorithm which computes the LLL reduction does not do permutations as Algorithm \ref{a:LLL} does,
e.g., the algorithm permutes two columns which are not consecutive or permutes two consecutive columns
but the  corresponding Lov\'{a}sz condition is not satisfied after the permutation,
then we cannot guarantee this specific LLL reduction will increase $P_\sB$.

It is interesting to note that  \cite{LingH07} showed that all the size reductions on the off-diagonal
entries above the  superdiagonal  of $\R$ have no effect on the residual norm of the Babai  point.
Here we see  that  those size reductions are not useful from another perspective.

If we do not do size reductions in Algorithm \ref{a:LLL}, the algorithm will do only column
permutations. We refer to this column permutation strategy as LLL-permute.
The  column permutation strategies SQRD \cite{WBRKK} and V-BLAST \cite{FGVW}
are often used for solving box-constrained ILS problems (see \cite{DamGC03} and \cite{ChaH08}).
In the following, we give simple numerical test results to see how the four methods
(SQRD, V-BLAST, LLL-permute with $\delta=1$ and LLL with $\delta=1$) affect  $P_\sB$.

We performed our \textsc{Matlab} simulations for the following two cases.

\begin{itemize}
\item  Case 1. $\A=\text{randn}(n,n)$,
where $\text{randn}(n,n)$ is a \textsc{Matlab} built-in function
to  generate a random $n\times n$ matrix, whose entries follow the normal distribution ${\cal N}(0,1)$.

\item  Case 2. $\A=\U\D\V^T$, $\U,\V$ are random orthogonal matrices obtained by the QR factorization of random matrices generated by
$\text{randn}(n,n)$ and $\D$ is a $n\times n$ diagonal matrix with $d_{ii}=10^{3(n/2-i)/(n-1)}$.

\end{itemize}

\begin{figure}[!htbp]
\centering
\includegraphics[width=3.2in]{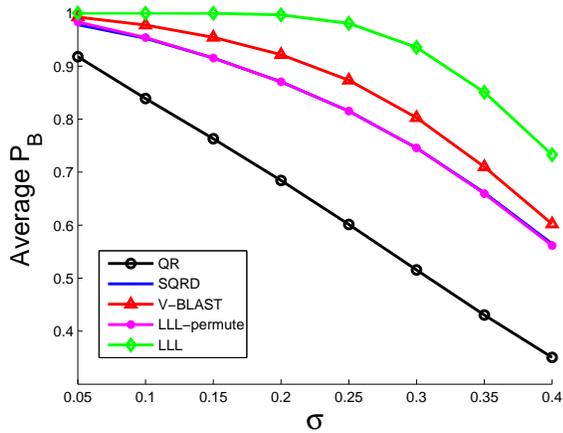}
\caption{Average success probability  versus $\sigma$ for Case 1, $n=20$} \label{fig:LLLSQVB1}
\end{figure}

\begin{figure}[!htbp]
\centering
\includegraphics[width=3.2in]{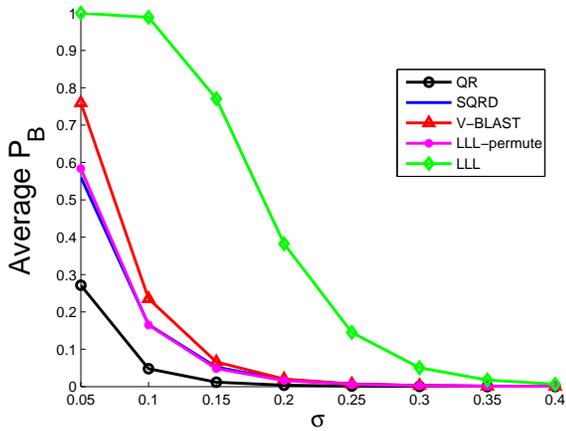}
\caption{Average success probability versus $\sigma$ for Case 2, $n=20$} \label{fig:LLLSQVB2}
\end{figure}

In the tests for each case for a fixed $n$ we gave 200 runs to generate 200 different $\A$'s.
For $n=20$, Figures  \ref{fig:LLLSQVB1} and \ref{fig:LLLSQVB2}
display  the average success probabilities
of the Babai points corresponding to various reduction or permutation strategies over 200 runs versus $\sigma=0.05:0.05:0.4$,
for Cases 1 and 2, respectively. In both figures, ``QR'' means the QR factorization is used, giving $\Pr(\x^\sB=\hbx)$.

From Figures \ref{fig:LLLSQVB1} and \ref{fig:LLLSQVB2}, we can see that on average the LLL reduction
improves $P_\sB$ much more significantly than the other three,
V-BLAST performs better than LLL-permute and SQRD, and LLL-permute and SQRD have similar performance.
We observed the same phenomenon when we changed the dimensions of $\A$.

Figures \ref{fig:LLLSQVB1} and \ref{fig:LLLSQVB2} indicate that on average SQRD and V-BLAST increase $P_\sB$.
However, unlike LLL-permute, both SQRD and V-BLAST may decrease $P_\sB$ sometimes.
Table~\ref{tb:LLLSQVB}  gives the number of runs out of 200 in which SQRD and V-BLAST decrease $P_\sB$
for various $\sigma$ and $n$.
From the table we can see that for both Cases 1 and 2, the chance that SQRD decreases $P_\sB$ is much larger than V-BLAST and when $\sigma$ increases, the chance that SQRD decreases $P_\sB$ tends to decrease.
For Case 2, when $n$ increases, the chance that SQRD decreases $P_\sB$ tends to decrease, but this phenomenon is not seen for Case 1.

\begin{table}[h!]
\caption{Number of runs out of 200 in which $P_\sB$ decreases}
\vspace*{-2mm}
\centering
\begin{tabular}{|c||c||c|c|c||c|c|c|}
 \hline
 \multicolumn{2}{|c||}{}  &  \multicolumn{3}{c||}{Case 1} & \multicolumn{3}{c|}{Case 2} \\ \hline
Methods  &  \backslashbox{$n$}{$\sigma$} & $0.1$ & $0.2$ & $0.3$ &  $0.1$ & $0.2$ & $0.3$ \\ \hline
  & 10  & 9 &10 &6 &13  & 8 &5 \\   %\cline{2-8}
SQRD & 20  & 12 &11 &7 &6  & 2 &1 \\  % \cline{2-8}
  & 30  & 16 &14 & 11& 0 &1  & 1\\  % \cline{2-8}
  & 40  & 15 &9 & 5& 0 & 0 &0 \\  \hline
  & 10  & 0 &0 &0 & 2 & 6 &7 \\   %\cline{2-8}
V-BLAST & 20  & 0 &0 &0 & 0 & 0 &0 \\  %\cline{2-8}
  & 30  & 0 &0 & 0&0  & 0 &0 \\  % \cline{2-8}
 & 40  & 0 & 0& 0&0  &0  & 0\\ \hline
\end{tabular}
\label{tb:LLLSQVB}
\end{table}

\vspace*{-8mm}

%%%%%%%%%%%%%%%%%%%%%%%%%%%%%%%%%%%%%%%%%%%%%%
\subsection{Effects of $\delta$ on the enhancement of $P_\sB$ }\label{s:delta}
Suppose  that $\R_1$ and $ \R_2$ are obtained by applying Algorithm \ref{a:LLL} to $\A$
with $\delta=\delta_1$ and $\delta=\delta_2$, respectively and $\delta_1 < \delta_2$.
A natural question is what is the relation between $P_\sB(\R_1)$ and $P_\sB(\R_2)$?
In the following we try to address this question.
First we give a result for $n=2$.

\begin{theorem}\label{t:2delta}
Suppose  that $\R_1$ and $ \R_2$ are obtained by applying Algorithm \ref{a:LLL} to $\A\in \mathbb{R}^{m\times n}$
with $\delta=\delta_1$ and $\delta=\delta_2$, respectively and $\delta_1 < \delta_2$.
If $n=2$,  then
\beq
P_\sB(\R_1)\leq P_\sB(\R_2). \label{eq:p12}
\eeq
 \end{theorem}

{\em Proof}.
Note that only two columns are involved in the reduction process and the value of $\delta$ only determines
when the process should terminate.
In the reduction process, the upper triangular matrix $\R$ either first becomes $\delta_1$-LLL reduced and then becomes $\delta_2$-LLL reduced
after some more permutations or  becomes $\delta_1$-LLL reduced and $\delta_2$-LLL reduced at the same time.
Therefore, by Lemma \ref{l:probper} the conclusion holds.
\ \ $\Box$

\medskip

However, the inequality \eqref{eq:p12} in Theorem \ref{t:2delta} may not hold when $n\geq3$.
In fact, for any given $n\geq3$, we can give an   example to illustrate this.

%\medskip

\bex \label{1delta}
Let $\delta_1$ and $\delta_2$ satisfy $1/4<\delta_1 < \delta_2\leq1$ and $\delta_2<\delta_1^2+1/4$.
Let $\eta$ and $\theta$ satisfy $\delta_1 <\eta< \delta_2$ and
$0 < \theta < \frac{1}{2}\sqrt{\delta_1(\eta-\delta_1)}$.
Let
\beq
\R=\bmx    1  & 0 &   1/2\\
         0 &   \sqrt{\eta}   & \theta\\
         0    &     0  &  \delta_1
    \emx.
\label{eq:ar}
\eeq
Note that $\R$ is size reduced already.

Suppose that we apply Algorithm \ref{a:LLL} with $\delta=\delta_1$ to $\R$, leading to $\R_1$.
The  first two columns of $\R$ do not  permute as the Lov\'{a}sz condition holds.
However, the Lov\'{a}sz condition does not hold for the last two columns
and a permutation is needed.
Then by Lemma \ref{l:probper} we must have $P_\sB(\R_1)> P_\sB(\R)$.

Applying Algorithm \ref{a:LLL} with $\delta=\delta_2$ to $\R$,  we obtain
$$
\R_2=\bmx    \sqrt{\eta}   & 0 &   \theta\\
         0 &   1 & 1/2\\
         0    &     0 & \delta_1
    \emx,
$$
whose diagonal entries are the same as those of $\R$ with a different order.
Then we have $P_\sB(\R_2)=P_\sB(\R)$.
Therefore, $P_\sB(\R_1) > P_\sB(\R_2)$.

With  $\R \in \mathbb{R}^{3\times 3}$ given in \eqref{eq:ar},
we define $\A$ as $\A=\bsmx    \R   &    0\\
         0    &     \I_{n-3}    \esmx \in \Rnbn$,
it is easy to show that we still have $P_\sB(\R_1)>P_\sB(\R_2)$, where $\R_1$ and $\R_2$ were obtained by
applying Algorithm \ref{a:LLL} to $\A$ with $\delta=\delta_1$ and $\delta=\delta_2$, respectively.
\eex
\medskip

Although the above example shows that larger $\delta$ may not guarantee to produce higher $P_\sB$
when $n\geq 3$,  we can expect that the chance that $P_\sB(\R_1) \leq P_\sB(\R_2)$ is much higher than the chance
that $P_\sB(\R_1) > P_\sB(\R_2)$.
Here we give an explanation.
If $\R_1$ is not $\delta_2$-LLL reduced, applying Algorithm \ref{a:LLL} with $\delta=\delta_2$ to $\R_1$
produces $\bbR_1$ with $P_\sB(\bbR_1) \geq P_\sB(\R_1)$.
Although $\bbR_1$ may not be equal to $\R_2$, we can expect that the difference between
these two $\delta_2$-LLL reduced matrices is small.
Thus it is likely that $P_\sB(\R_2) \approx P_\sB(\bbR_1)\geq P_\sB(\R_1)$.

\begin{center}
\begin{figure}[h!]
%\centering
\includegraphics[width=3.3in]{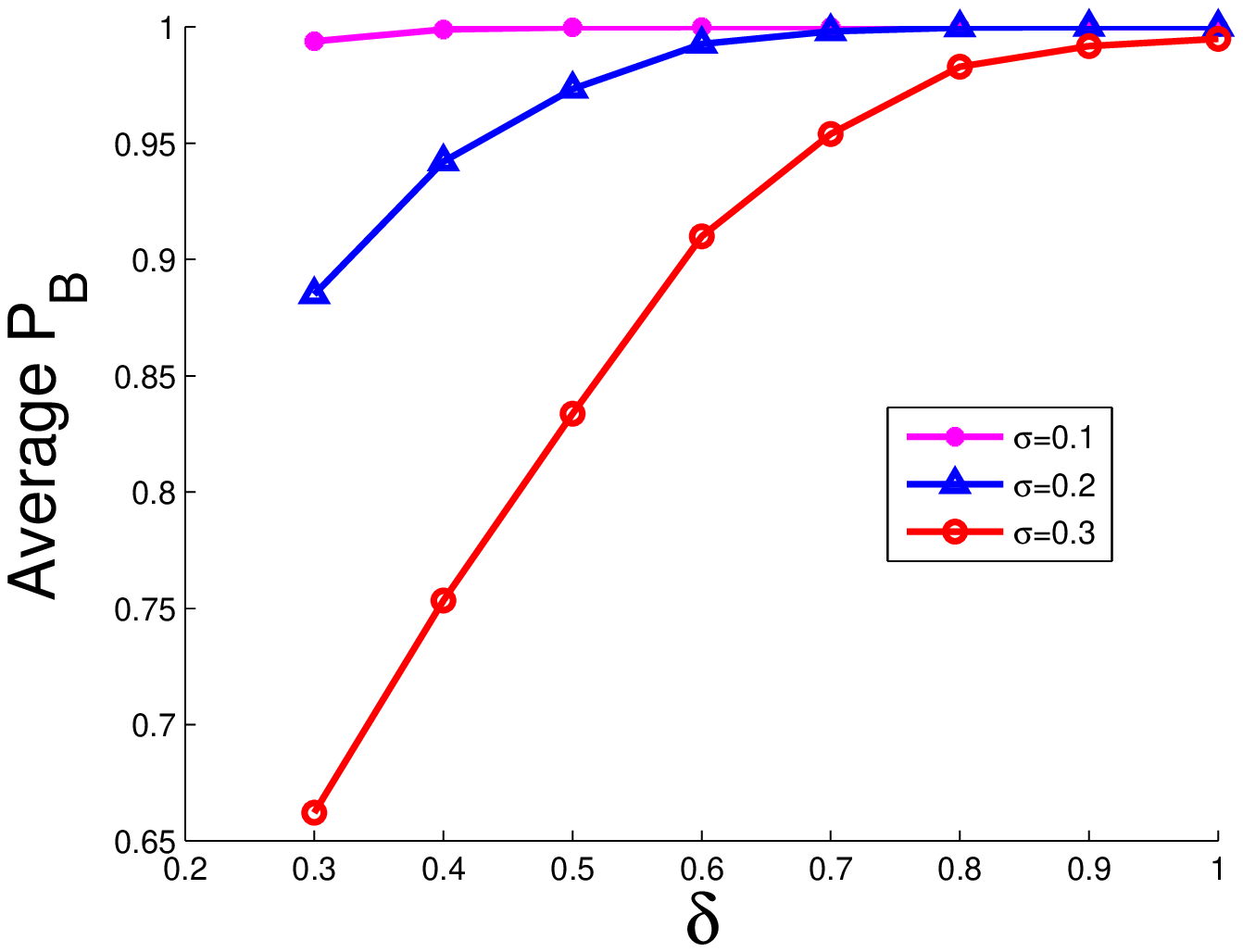}
\caption{Average $P_\sB$ after the LLL reduction for Case 1, $n=20$} \label{fig:delta1}
\end{figure}
\end{center}

\begin{center}
\begin{figure}[h!]
%\centering
\includegraphics[width=3.3in]{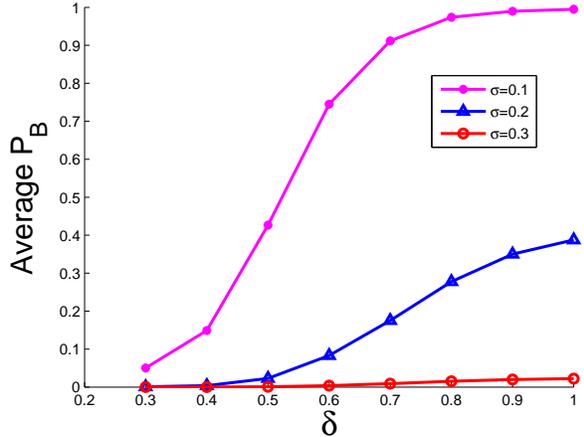}
\caption{Average $P_\sB$ after the LLL reduction  for Case 2, $n=20$} \label{fig:delta2}
\end{figure}
\end{center}

Here we give numerical results to show how $\delta$ affects $P_\sB$ (i.e., $\Pr(\z^B=\hat{\z})$).
We used the  matrices defined in Cases 1 and 2 of Section \ref{s:lllprob}.
As before, in the tests for each case we gave 200 runs to generate 200 different $\A$'s for a fixed $n$.
For $n=20$, Figures \ref{fig:delta1} and \ref{fig:delta2} display  the average $\Pr(\z^B=\hat{\z})$ over 200 runs versus
$\delta=0.3\!:\!0.1\!:\!1.0$  for Cases 1 and 2, respectively.
The three curves in both figures  correspond to $\sigma=0.1, 0.2, 0.3$.
For comparisons, we give the corresponding $\Pr(\x^B=\hat{\x})$ in the following table.
\vspace*{-1mm}
\begin{table}[h!]
\caption{Success probability $\Pr(\x^B=\hat{\x})$}
\vspace*{-1mm}
\centering
\begin{tabular}{|c||c|c|c|}
\hline
 & $\sigma=0.1$ & $\sigma=0.2$ & $\sigma=0.3$ \\  \hline
Case 1 & 0.839 & 0.661 & 0.477  \\  \hline
Case 2 & $1.85\times 10^{-2}$ &  $1.95 \times 10^{-4}$ &  $5.56 \times 10^{-6}$ \\ \hline
\end{tabular}
\label{tb:qr}
\end{table}

From Table \ref{tb:qr}, Figures \ref{fig:delta1} and \ref{fig:delta2}, we can see that the LLL reduction has a significant effect
on improving $P_\sB$.
Figures \ref{fig:delta1} and \ref{fig:delta2} show that as $\delta$ increases, on average $P_\sB$ increases too,
in particular for large $\sigma$.
But we want to point out that we also noticed that sometimes a larger $\delta$   resulted in a smaller $P_\sB$ in the  tests.
Table  \ref{tb:delta}   gives  the exact number of runs out of those 200 runs
in which $P_\sB$ decreases when $\delta$ increases from $t$ to $t+0.1$ for $t=0.3:0.1:0.9$.
From Table  \ref{tb:delta}  we can see that most of the time $P_\sB$ does not decrease when $\delta$ increases.
We would like to point out that in our numerical tests we tried various dimension size $n$
for the two test cases and observed the same phenomena.

\begin{table}[h!]
\caption{Number of runs  in which $P_\sB$ decreases when $\delta$ increases}
\centering
\begin{tabular}{|c||c|c|c||c|c|c|}
\hline
  & \multicolumn{3}{c||}{Case 1} & \multicolumn{3}{c|}{Case 2} \\  \hline
%  & \multicolumn{3}{c||}{$\sigma$} & \multicolumn{3}{c|}{$\sigma$} \\ \cline{2-7}
\backslashbox{$\delta$}{$\sigma$}& $0.1$ & $0.2$ & $0.3$ &  $0.1$ & $0.2$ & $0.3$ \\ \hline
0.3---0.4 & 8  & 9  & 10 & 9 & 10 & 11  \\
0.4---0.5 & 10 & 9  & 8  & 10 & 11 & 11  \\
0.5---0.6 & 13 & 14 & 13 & 12 & 11 & 11 \\
0.6---0.7 & 19 & 18 & 16 & 17 & 18 & 20  \\
0.7---0.8 & 2  & 10 & 12 & 12 & 13 & 14  \\
0.8---0.9 & 3  & 11 & 9  & 15 & 18 & 19  \\
0.9---1.0 & 1  & 13 & 8  & 16 & 19 & 22  \\ \hline
\end{tabular}
\label{tb:delta}
\end{table}

\vspace*{-10mm}

\subsection{Some upper bounds on $P_\sB$ after the LLL reduction}\label{s:ub}
We have shown that the LLL reduction by Algorithm \ref{a:LLL} can enhance the success probability of the Babai point.
A natural question is how much is the enhancement?
If the LLL reduction has been computed by Algorithm \ref{a:LLL},
then we can easily obtain the ratio $\Pr(\z^\sB=\hbz)/\Pr(\x^\sB=\hbx)$ by using the formula given in \eqref{e:pb}.
If we  only know the R-factor of the QR factorization of $\A$, usually it is impossible
to know the ratio exactly.
However, we will derive some bounds on $\Pr(\z^\sB=\hbz)$, which involve only the R-factor of the QR factorization of $\A$.
From these bounds one can immediately obtain bounds on the ratio.

Before giving an upper bound on $\Pr(\z^\sB=\hbz)$, we  give the following result,
see, e.g., \cite[Thm 6]{NguS09}.

\begin{lemma}\label{l:lub}
Let $\R$ be the R-factor of the QR factorization of $\A$
and let $\R^{(p)}$ be the upper triangular matrix after the $p$-th column permutation and triangularization in
the LLL reduction process by Algorithm \ref{a:LLL}, then
for $i=1,2,\ldots,n$
\beq \label{in:lub}
\begin{split}
  &  \min\{r_{ii}, r_{i+1,i+1}, \ldots, r_{nn}\}   \\
 \leq  & \,  r_{ii}^{(p)}      \leq \max\{r_{11},r_{22},\ldots, r_{ii}\}.
\end{split}
\eeq
 \end{lemma}

When the LLL reduction process finishes, the diagonal entries of the  upper triangular matrix
certainly satisfy \eqref{in:lub}.
Then using the second inequality in \eqref{in:lub} we obtain the following result from  \eqref{e:pb}.

\begin{theorem}\label{t:probub1}
Suppose that the ILS problem \eqref{e:ILSR} is transformed to the ILS problem \eqref{e:reduced}
after the LLL reduction by Algorithm \ref{a:LLL}.
The success probability of the Babai point for the ILS problem \eqref{e:reduced} satisfies:
\beq
\Pr(\z^\sB=\hbz)\leq \Pi_{i=1}^n\phi(\gamma_i),  \label{e:probub1}
\eeq
where $\gamma_i=\max\{r_{11},r_{22},\cdots, r_{ii}\}$.
 \end{theorem}

In the following we give another upper bound on the success probability of the Babai point,
which is invariant to the unimodular transformation to $\R$.
The result was  essentially obtained in \cite{Teun03}, but our proof is much simpler.
\begin{lemma}\label{l:upp}
Let $\R \in \Rnbn$ be an upper triangular matrix with positive diagonal entries, then
\beq \label{in:uppphi}
\prod_{i=1}^n\phi(r_{ii}) \leq \phi^n\Big(\Big(\prod_{i=1}^n r_{ii}\Big)^{1/n}\Big),
\eeq
where the equality holds if and only if all the diagonal entries of $\R$ are equal.
\end{lemma}

{\em Proof}. Let $h(\xi)=\ln(\phi(\exp(\xi))$ and $v_i=\ln r_{ii}$ for $i=1,\ldots, n$.
Define $v= \frac{1}{n} \sum_{i=1}^n v_i = \frac{1}{n}\ln(\prod_{i=1}^n r_{ii})$.
To prove \eqref{in:uppphi}, it suffices to show that
\beq \label{in:uppF}
 \frac{1}{n} \sum_{i=1}^n h(v_i) \leq h(v).
\eeq

It is easy to verify that
$$
h''(\xi)=\frac{1}{2\sigma}\exp(\xi) g'\Big(\frac{1}{2\sigma}\exp(\xi)\Big),
$$
where $g(\cdot)$ was defined in the proof of Lemma \ref{l:probper}.
According to   the proof of Lemma \ref{l:probper}, $g'(\zeta)<0$ for $\zeta>0$.
Thus  $h''(\xi)<0$, i.e.,  $h(\xi)$ is a strictly concave function.
Therefore,  \eqref{in:uppF} must hold and the equality holds if and only if all $v_i$
are equal, or equivalently all $r_{ii}$ are equal.
\ \ $\Box$

Suppose that the ILS problem \eqref{e:ILSR} is transformed to the ILS problem \eqref{e:reduced}
after the LLL reduction by Algorithm \ref{a:LLL}.
Then $\det(\bbR)= \det(\R)=\prod_{i=1}^n r_{ii}$.
Thus by Lemma \ref{l:upp} we have
\beq
\Pr(\z^\sB=\hbz)=\prod_{i=1}^n\phi(\br_{ii})
\leq      \phi^n \Big(\Big(\prod_{i=1}^n r_{ii}\Big)^{1/n}\Big)  \label{e:probub2} % \phi^n({\det}^{1/n}(\R)).
\eeq
The upper bound  is reachable if and only if
all the diagonal entries of $\bbR$  are equal to ${\det}^{1/n}(\R)$.
If the gap between the largest diagonal entry and the smallest diagonal entry of $\bbR$ is large,
the upper bound in \eqref{e:probub2} will not be tight.
In the following, we give an improved upper bound.

\begin{theorem}\label{t:upp}
Under the same assumption as in Theorem \ref{t:probub1},
if there exist   indices $i_1, i_2, \ldots, i_l$ such that
\beq
M_k \leq m_{k+1}, \quad k=1, \ldots, l, \label{in:upp}
\eeq
where
\begin{align*}
M_k & = \max\{r_{i_{k-1}+1,i_{k-1}+1}, r_{i_{k-1}+2,i_{k-1}+2}, \ldots, r_{i_{k},i_{k}}\}  \nonumber   \\
m_{k+1}  &= \min\{r_{i_{k}+1,i_{k}+1}, r_{i_{k}+2,i_{k}+2},\ldots, r_{i_{k+1},i_{k+1}}\},
\end{align*}
with $i_0=0$ and $i_{l+1}=n$,
then
\beq
\Pr(\z^\sB=\hbz)
 \leq \prod_{k=1}^{l+1} \phi^{i_{k}-i_{k-1}} (\nu_k)
\leq \phi^n(\nu),   \label{e:probub3}    %  \label{e:probub4}
\eeq
where
$$
\nu_k= \biggl(\prod_{j=i_{k-1}+1}^{i_{k}}r_{jj}\biggr)^{1/(i_{k}-i_{k-1})}, \quad
\nu=\biggl(\prod_{j=1}^{n}r_{jj}\biggr)^{1/n}.
$$
 \end{theorem}

{\em Proof}. Partition $\R$  as follows:
$$
\R=[\R_1,\R_2,\cdots, \R_{l+1}],
$$
where the diagonal entries of $\R$ which are in block $\R_k\in \mathbb{R}^{n\times (i_k-i_{k-1})}$
are $r_{i_{k-1}+1,i_{k-1}+1}$, $r_{i_{k-1}+2,i_{k-1}+2}$,
$\ldots$,  $r_{i_{k},i_{k}}$ for $k=1, \ldots, l+1$.
The condition \eqref{in:upp} is to ensure that in the LLL reduction process by Algorithm \ref{a:LLL}
there are no column permutations between $\R_k$s.
Now we prove this claim.
Suppose that Algorithm \ref{a:LLL} has just finished the operations on $\R_2$ and is going to work on $\R_3$.
At this moment, $[\R_1,\R_2]$ is LLL reduced.
In the LLL reduction of $[\R_1,\R_2]$, no column permutation between the last column of $\R_1$ and and the first column
of $\R_2$ occurred.
In fact,  by \eqref{in:lub} in Lemma \ref{l:lub} and the  inequality $M_1 \leq m_2$ from \eqref{in:upp},
after a permutation, say the $p$-th permutation,  in the LLL reduction of  $[\R_1,\R_2]$ by Algorithm \ref{a:LLL},
\begin{align*}
 r_{i_1,i_1}^{(p)} & \leq  \max\{r_{11}, \ldots, r_{i_1,i_1}\} \\
& \leq    \min\{r_{i_1+1,i_1+1},\cdots, r_{i_2,i_2}\} \leq r_{i_1+1,r_1+1}^{(p)}.
\end{align*}
Thus for any $\delta$ satisfying $1/4< \delta\leq1$, the Lov\'{a}sz condition \eqref{e:criteria2} is satisfied for columns $i_1$ and $i_1+1$
and no permutation between these two columns would occur.
Now the algorithm goes to work on the first column of $\R_3$.
Again we can similarly show that no column permutation between the last column of $\R_2$ and and the first column
of $\R_3$ will occur, so the algorithm will not go back to $\R_2$.
The algorithm continues and whenever the current block is LLL reduced it goes to next block and will not come
back to the previous block.
Then by applying the result given in \eqref{e:probub2} for each block $\R_k$ we obtain the first inequality in \eqref{e:probub3}.
The second inequality in \eqref{e:probub3} is obtained immediately by applying Lemma \ref{l:upp}. \  \ $\Box$
\medskip

If  indices $i_k$ for $k=1,\ldots, l$ defined in  Theorem \ref{t:upp}
do not exist, we assume $l=0$, then the first inequality in \eqref{e:probub3}
still holds as its right hand side is just $\phi^n(\nu)$.

We now show how to find these indices if they exist.
It is easy to verify that \eqref{in:upp} is equivalent to
\beq
\max\{M_1, \ldots, M_k\} \leq \min \{m_{k+1}, \ldots, m_{l+1}\}
\label{e:Mm}
\eeq
for $k=1,\ldots, l$.
Define two vectors $\u, \v \in \mathbb{R}^{n-1}$  as follows:
$u_1=r_{11}$, $u_i=\max\{r_{11}, \ldots, r_{ii}\}=\max\{u_{i-1}, r_{ii}\}$
for $i=2,\ldots, n-1$;
$v_{n-1}=r_{nn}$, $v_i=\min\{r_{i+1,i+1}, \ldots, r_{nn}\}=\min\{r_{i+1,i+1}, v_{i+1}\}$.
Then \eqref{e:Mm} is equivalent to
$$
u_{i_k} \leq v_{i_k}, \quad k=1, \ldots, l.
$$
Thus we can compare the entries of $\u$ and $\v$ from the first to the last to obtain all indices $i_k$.
It is easy to observe that that the total cost is $O(n)$.

Let $\beta_1$, $\beta_2$ and $\beta_3$  denote the three upper bounds on $\Pr(\z^\sB=\hbz)$ given
in \eqref{e:probub1} and \eqref{e:probub3}, respectively, i.e.,
$$
\beta_1= \Pi_{i=1}^n\phi(\gamma_i), \ \
\beta_2=\prod_{k=1}^{l+1} \phi^{i_{k}-i_{k-1}} (\nu_k), \ \
\beta_3= \phi^n(\nu).
$$
In the following, we first give some special examples to compare $\beta_1$, $\beta_2$ and $\beta_3$.

\bex \label{ex:ub2}
Let $\R= \bmx 1/\eta & \times \\ 0 & \eta^2 \emx$, where $0<\eta<1$ and $\times$ is any real number.
Then
$$
\beta_1=\phi^2(1/\eta), \ \ \beta_2=\beta_3 = \phi^2(\sqrt{\eta}).
$$
By the definition of $\phi(\zeta)$ given in \eqref{e:pb},  $\phi(1/\eta) \rightarrow 1$  and $\phi(\sqrt{\eta})\rightarrow 0$
when $\eta \rightarrow 0$.
Thus,  when  $\eta$ is very small, $\beta_2$ and $\beta_3$ are much sharper than $\beta_1$.
\eex

\bex \label{ex:ub1}
Let
$$
\R=\bmx
          \eta/3&  \times &  \times  &  \times \\
         0 &\eta & \times  &  \times \\
         0& 0& 1/\eta^3 & \times  \\
           0& 0& 0& \eta/2
    \emx, \quad 0<\eta<1,
$$
where $\times$ is any real number.
Then
\begin{align*}
& \beta_1 = \phi(\eta/3)\phi(\eta)\phi^2(1/\eta^3), \\
& \beta_2  =  \phi(\eta/3) \phi^3\big(\sqrt[3]{1/(2\eta)}\big), \ \
\beta_3 = \phi^4 (\sqrt[4]{1/6}).
\end{align*}
From the definition of $\phi(\zeta)$, we see that when $\eta\rightarrow 0$,
$$
\beta_1  \rightarrow 0, \ \ \beta_2  \rightarrow 0,  \ \
\beta_1 /\beta_2  \rightarrow 0, \ \ \beta_2 /\beta_3  \rightarrow 0.
$$
Therefore, when $\eta$ is very small,  $\beta_1$ is much sharper than
  $\beta_2$, which is also much sharper than  $\beta_3 $.
\eex

\medskip

Now we use more general examples to compare the three upper bounds and also compare them with  $\Pr(\z^\sB=\hbz)$.
In additional to Cases 1 and 2 given in Section \ref{s:lllprob}, we also tested the following case:

Case 3. $\A=\Q\R$, where $\Q$ is a random orthogonal matrix obtained by the QR factorization of a random matrix generated by
$\text{randn}(n,n)$ and $\R$ is an $n\times n$ upper triangular matrix with $r_{ii}^2$ following the $\chi^2$ distribution with
 freedom degree $i$ and with $r_{ij}$ ($j>i$) following the normal distribution ${\cal N}(0,1)$.

Case 3 is motivated by Case 1. In Case 1,  the entries of the R-factor of the QR factorization of $\A$
have the same distributions as the entries of $\R$ in Case 3, except that the freedom degree for $r_{ii}^2$
is $n-i+1$, see \cite[p99]{Mui82}.

In the numerical experiments, for a given $n$ and for each case, we gave 200 runs to generate 200 different  $\A$'s.

All the six tables given below display the average values of $\Pr(\x^\sB=\hbx)$ (corresponding to QR),
$\Pr(\z^\sB=\hbz)$ (corresponding to LLL with $\delta=1$),
$\beta_1$, $\beta_2$ and $\beta_3$.
For each case, we give two tables.
In the first table, $n$ is fixed and $\sigma$ varies,
and in the second table, $n$ varies and $\sigma$ is fixed.
In Tables \ref{tb:bd21} and \ref{tb:bd23} $\sigma$ was fixed to be 0.4, while
in Table  \ref{tb:bd22} $\sigma$ was fixed to be 0.1.
We used different values of $\sigma$ for these three tables so that $\Pr(\z^\sB=\hbz)$
is neither close to 0 nor close to 1, otherwise the bounds would not be much interesting.

For Case 1, from Tables \ref{tb:bd11} and \ref{tb:bd21} we observe that the upper bounds
$\beta_2$ and $\beta_3$ are sharper than the upper bound $\beta_1$, especially when $n$ is small,
and  the former are good approximations to $\Pr(\z^\sB=\hbz)$.

For Case 2, from Table \ref{tb:bd12} we observe that  the upper bound $\beta_1$ is extremely loose
when $\sigma$ is large, and  $\beta_2$ and $\beta_3$ are much sharper for all those $\sigma$.
From Table \ref{tb:bd22} we see that when $n$ becomes larger, the upper bounds $\beta_2$ and $\beta_3$
become worse, although they  are still sharper than $\beta_1$.
Tables \ref{tb:bd12}-\ref{tb:bd22} show that  $\beta_2$ is equal to $\beta_3$.
Actually it is indeed true.

For Case 3, from Tables \ref{tb:bd13} and \ref{tb:bd23} we observe that the success probability of the Babai point
improves after the LLL reduction, but not as much as Cases 1 and 2.
We also observe that $\beta_2$ is sharper than $\beta_1$, both are much sharper than $\beta_3$,
and $\beta_2$ is a reasonable approximation to $\Pr(\z^\sB=\hbz)$.

Based on the  numerical experiments and Theorem \ref{t:upp} we suggest taking $\min\{\beta_1,\beta_2\}$
as an upper bound on $\Pr(\z^\sB = \hbz)$ in practice.

Although  the upper bound $\min\{\beta_1,\beta_2\}$ is a good approximation to $\Pr(\z^\sB=\hbz)$
in the above numerical tests,  we want to point out that this upper bound can be very loose.
Here is a contrived  example:  Suppose all the off-diagonal entries of $\R$  in  Example \ref{ex:ub1} are zero.
Then
$$
\Pr(\x^\sB=\hbx)\!=\!\Pr(\z^\sB=\hbz)\!=\! \phi(\eta/3) \phi(\eta)\phi(1/\eta^3)\phi(\eta/2).
$$
Thus, when $\eta \rightarrow 0$,
$\Pr(\z^\sB=\hbz)/\min\{\beta_1,\beta_2\}  \rightarrow 0.$

\begin{table}[b!]
\caption{Average $P_\sB$ and bounds for Case 1, $n=20$}
\centering
\begin{tabular}{|c||c|c|c|c|c|}
\hline
$\sigma$ & QR & LLL & $\beta_1$ & $\beta_2$ & $\beta_3$ \\ \hline
0.05 & 0.93242 & 1.00000 & 1.00000 & 1.00000 & 1.00000 \\ \hline
0.10 & 0.84706 & 1.00000 & 1.00000 & 1.00000 & 1.00000 \\ \hline
0.15 & 0.75362 & 0.99999 & 1.00000 & 1.00000 & 1.00000 \\ \hline
0.20 & 0.66027 & 0.99966 & 1.00000 & 0.99984 & 0.99984 \\ \hline
0.25 & 0.56905 & 0.99815 & 1.00000 & 0.99891 & 0.99891 \\ \hline
0.30 & 0.48130 & 0.99289 & 1.00000 & 0.99645 & 0.99645 \\ \hline
0.35 & 0.39864 & 0.97589 & 0.99999 & 0.98849 & 0.98849 \\ \hline
0.40 & 0.32279 & 0.93432 & 0.99997 & 0.96319 & 0.96319 \\ \hline
\end{tabular}
\label{tb:bd11}
\end{table}

\begin{table}[h!]
\caption{Average $P_\sB$ and bounds for Case 1, $\sigma=0.4$}
\centering
\begin{tabular}{|c||c|c|c|c|c|}
\hline
$n$ & QR & LLL & $\beta_1$ & $\beta_2$ & $\beta_3$ \\ \hline
5 & 0.37181 & 0.52120 & 0.92083 & 0.55777 & 0.56437 \\ \hline
10 & 0.33269 & 0.73310 & 0.99634 & 0.75146 & 0.75146 \\ \hline
15 & 0.30324 & 0.87116 & 0.99967 & 0.89076 & 0.89076 \\ \hline
20 & 0.32896 & 0.94211 & 0.99999 & 0.97004 & 0.97004 \\ \hline
25 & 0.31439 & 0.95364 & 1.00000 & 0.98993 & 0.98993 \\ \hline
30 & 0.32649 & 0.96961 & 1.00000 & 0.99752 & 0.99752 \\ \hline
35 & 0.34107 & 0.97361 & 1.00000 & 0.99939 & 0.99939 \\ \hline
40 & 0.32538 & 0.97579 & 1.00000 & 0.99980 & 0.99980 \\ \hline
\end{tabular}
\label{tb:bd21}
\end{table}

\begin{table}[h!]
\caption{Average $P_\sB$ and bounds for Case 2, $n=20$}
\centering
\begin{tabular}{|c||c|c|c|c|c|}
\hline
$\sigma$ & QR & LLL & $\beta_1$ & $\beta_2$ & $\beta_3$ \\ \hline
0.05 & 0.27379 & 1.00000 & 1.00000 & 1.00000 & 1.00000 \\ \hline
0.10 & 0.01864 & 0.99490 & 1.00000 & 0.99939 & 0.99939 \\ \hline
0.15 & 0.00161 & 0.82023 & 1.00000 & 0.89650 & 0.89650 \\ \hline
0.20 & 0.00019 & 0.38963 & 1.00000 & 0.46930 & 0.46930 \\ \hline
0.25 & 0.00003 & 0.10896 & 1.00000 & 0.13462 & 0.13462 \\ \hline
0.30 & 0.00001 & 0.02248 & 1.00000 & 0.02738 & 0.02738 \\ \hline
0.35 & 0.00000 & 0.00411 & 1.00000 & 0.00489 & 0.00489 \\ \hline
0.40 & 0.00000 & 0.00074 & 1.00000 & 0.00086 & 0.00086 \\ \hline
\end{tabular}
\label{tb:bd12}
\end{table}

\begin{table}[h!]
\caption{Average $P_\sB$ and bounds for Case 2, $\sigma=0.1$}
\centering
\begin{tabular}{|c||c|c|c|c|c|}
\hline
$n$ & QR & LLL & $\beta_1$ & $\beta_2$ & $\beta_3$ \\ \hline
5 & 0.06157 & 0.75079 & 0.99984 & 0.83688 & 0.83688 \\ \hline
10 & 0.05522 & 0.98875 & 1.00000 & 0.99344 & 0.99344 \\ \hline
15 & 0.03069 & 0.99670 & 1.00000 & 0.99860 & 0.99860 \\ \hline
20 & 0.01865 & 0.99486 & 1.00000 & 0.99939 & 0.99939 \\ \hline
25 & 0.01149 & 0.97374 & 1.00000 & 0.99963 & 0.99963 \\ \hline
30 & 0.00562 & 0.88945 & 1.00000 & 0.99973 & 0.99973 \\ \hline
35 & 0.00324 & 0.76654 & 1.00000 & 0.99978 & 0.99978 \\ \hline
40 & 0.00175 & 0.68623 & 1.00000 & 0.99981 & 0.99981 \\ \hline
\end{tabular}
\label{tb:bd22}
\end{table}

\begin{table}[h!]
\caption{Average $P_\sB$ and bounds for Case 3, $n=20$}
\centering
\begin{tabular}{|c||c|c|c|c|c|}
\hline
$\sigma$ & QR & LLL & $\beta_1$ & $\beta_2$ & $\beta_3$ \\ \hline
0.05 & 0.91780 & 0.92401 & 0.92450 & 0.92471 & 1.00000 \\ \hline
0.10 & 0.85132 & 0.86372 & 0.87017 & 0.86856 & 1.00000 \\ \hline
0.15 & 0.77339 & 0.79087 & 0.80902 & 0.79945 & 1.00000 \\ \hline
0.20 & 0.68615 & 0.70836 & 0.74366 & 0.72379 & 1.00000 \\ \hline
0.25 & 0.59499 & 0.62040 & 0.67610 & 0.64530 & 0.99986 \\ \hline
0.30 & 0.50466 & 0.53153 & 0.60831 & 0.56704 & 0.99837 \\ \hline
0.35 & 0.41858 & 0.44528 & 0.54164 & 0.49161 & 0.99038 \\ \hline
0.40 & 0.33919 & 0.36432 & 0.47679 & 0.42031 & 0.96432 \\ \hline
\end{tabular}
\label{tb:bd13}
\end{table}

\begin{table}[h!]
\caption{Average $P_\sB$ and bounds for Case 3, $\sigma=0.4$}
\centering
\begin{tabular}{|c||c|c|c|c|c|}
\hline
5 & 0.35057 & 0.37086 & 0.47342 & 0.38878 & 0.53300 \\ \hline
10 & 0.35801 & 0.38542 & 0.49866 & 0.42252 & 0.75949 \\ \hline
15 & 0.32379 & 0.35068 & 0.47865 & 0.40583 & 0.90613 \\ \hline
20 & 0.34612 & 0.37149 & 0.49066 & 0.44551 & 0.96841 \\ \hline
25 & 0.35252 & 0.37865 & 0.48907 & 0.44248 & 0.99232 \\ \hline
30 & 0.32538 & 0.35542 & 0.46208 & 0.43224 & 0.99708 \\ \hline
35 & 0.33183 & 0.35421 & 0.46524 & 0.42288 & 0.99933 \\ \hline
40 & 0.32196 & 0.34759 & 0.45264 & 0.41220 & 0.99975 \\ \hline
\end{tabular}
\label{tb:bd23}
\end{table}

%%%%%%%%%%%%%%%%%%%%%%%%%%%%%%%%%%%%%%%%%%%%%%%%%%%%%
\section{Reduction of the search complexity by the LLL reduction}\label{s:comp}

In this section, we rigorously show that applying the LLL reduction algorithm given in Algorithm \ref{a:LLL}
can reduce the computational complexity of sphere decoders, which is measured approximately by the
number of nodes in the search tree.

%\subsection{Effects of LLL reduction on the search complexity }
The complexity  results of sphere decoders given in the literature are often about the complexity of enumerating all integer points
in the search region:
\beq
\label{e:search}
\|\tilde{\y}-\R\x\|_2 \leq \beta ,
\eeq
where $\beta$ is a constant called the search radius.
A typical measure of the complexity is the number of nodes enumerated  by  sphere decoders, which we denotes by $\zeta$.

For $i=n, n-1,\ldots,1$, define $E_i$ as follows
\beq
\label{e:beta}
E_i=\abs{\{\x_{i:n} \in \mbbZ^{n-i+1} :  \|\tby_{i:n}-\R_{i:n,i:n}\x_{i:n}\|_2 \leq \beta \}},
\eeq
where $|\cdot|$ denotes the number of elements in the set.
As given in \cite{PMW93}, $E_i $ can be estimated as follows:
\begin{equation}
E_i \approx\frac{V_{n-i+1}\,\beta^{n-i+1}}{\abs{\det(\R_{i:n,i:n})}}
=\frac{V_{n-i+1}\,\beta^{n-i+1}}{\abs{r_{ii}r_{i+1,i+1}\cdots r_{nn}}},
\label{e:Ei}
\end{equation}
where $V_{n-i+1}$ denotes the volume of an $(n-i+1)$-dimensional unit Euclidean ball.
%This estimation becomes the expected value to $E_i$ when $\x^\{RLS}$ is uniformly distributed over a Voroni cell of the lattice
This estimation would become the expected value to $E_i$ if $\tby_{i:n}$ is uniformly distributed over a Voroni cell of the lattice
generated by $\R_{i:n,i:n}$.
Then we have (see, e.g., \cite[Sec 3.2]{Abe11} and \cite{SeeJS11}).
\beq
\zeta=\sum_{i=1}^{n} E_i \approx \hat{\zeta}(\R)
\equiv \sum_{i=1}^{n}\frac{V_{n-i+1}\,\beta^{n-i+1}}{r_{ii}r_{i+1,i+1}\cdots r_{nn}}.
\label{e:complex}
\eeq
In practice, when  a sphere decoder such as the Schnorr-Euchner algorithm is used in the search process,
after an integer point is found, $\beta$ will be updated to shrink the search region.
But $\zeta$ or $\hat{\zeta}$ here does not take this into account for the sake of simplicity.

The following result shows that if the Lov\'{a}sz condition \eqref{e:criteria2} is not satisfied,
after a column permutation and triangularization, the  complexity $\hat{\zeta}(\R)$ decreases.

\begin{lemma}\label{t:compper}
Suppose that $\delta r_{k-1,k-1}^2 > r^2_{k-1,k}+r^2_{kk}$ for some $k$ for the $\R$ matrix in the ILS problem \eqref{e:ILSR}.
After the permutation of columns $k-1$ and $k$ and triangularization, $\R$ becomes $\bbR$,
i.e.,  $\bbR=\G_{k-1,k}^T\R \P_{k-1,k}$ {\rm (see \eqref{e:Permu})}.
Then the  complexity $\hat{\zeta}(\R)$ of the search process decreases  after the transformation, i.e.,
\beq
\hat{\zeta}(\R) > \hat{\zeta}(\bbR).
\label{e:compper}
\eeq
\end{lemma}

{\em Proof}. Since  $\bar{r}_{ii}=r_{ii}$ for $i\neq k-1, k$, \\$\br_{k-1,k-1}\br_{kk}=r_{k-1,k-1}r_{kk}$,
and $\br_{kk}>r_{kk}$, we have
\begin{align*}
&\hat{\zeta}(\R) - \hat{\zeta}(\bbR)\\
&= \sum_{i=1}^n \frac{V_{n-i+1}\,\beta^{n-i+1}}{r_{ii}r_{i+1,i+1}\cdots r_{nn}}
- \sum_{i=1}^n \frac{V_{n-i+1}\,\beta^{n-i+1}}{\bar{r}_{ii}\bar{r}_{i+1,i+1}\cdots \bar{r}_{nn}} \\
& =\frac{V_{n-k+1}\,\beta^{n-k+1}}{r_{kk}r_{k+1,k+1}\cdots r_{nn}}
- \frac{V_{n-k+1}\,\beta^{n-k+1}}{\bar{r}_{kk}r_{k+1,k+1}\cdots r_{nn}} \\
&= \left(\frac{1}{r_{kk}}-\frac{1}{\br_{kk}}\right)\frac{V_{n-k+1}\,\beta^{n-k+1}}{r_{k+1,k+1}\cdots r_{nn}}>0,
\end{align*}
completing the proof. \ \ $\Box$
\medskip

Suppose the Lov\'{a}sz condition \eqref{e:criteria2} does not hold for a specific $k$
and furthermore $|r_{k-1,k}|>r_{k-1,k-1}/2$.  The next lemma, which is analogous to Lemma \ref{l:probsr},
shows that the size reduction on $r_{k-1,k}$
performed before the permutation can decrease the complexity $\hat{\zeta}(\R)$ further.

\begin{lemma}\label{l:compsr}
 Suppose  that   in the ILS problem \eqref{e:ILSR} $\R$ satisfies
 $\delta r_{k-1,k-1}^2 > r^2_{k-1,k}+r^2_{kk}$ and $|r_{k-1,k}|>r_{k-1,k-1}/2$ for some $k$.
Let $\bbR$  be defined as in Lemma \ref{t:compper}.
Suppose  a size reduction on $r_{k-1,k}$ is performed first and then after the permutation of columns $k-1$ and $k$ and
 triangularization, $\R$ becomes $\hbR$, i.e.,  $\hbR=\hbG_{k-1,k}^T\R \Z_{k-1,k}\P_{k-1,k}$.
Then
\beq
\hat{\zeta}(\bbR) >  \hat{\zeta}(\hbR).
\label{e:compsr}
\eeq
\end{lemma}

{\em Proof}. By the same argument given in the proof of Lemma \ref{t:compper}, we have
$$
\hat{\zeta}(\bbR) - \hat{\zeta}(\hbR)
=\left(\frac{1}{\br_{kk}}-\frac{1}{\hr_{kk}}\right)\frac{V_{n-k+1}\,\beta^{n-k+1}}{r_{k+1,k+1}\cdots r_{nn}}.
$$
To show \eqref{e:compsr} we need only to prove $\br_{kk}<\hr_{kk}$.
Since $\br_{k-1,k-1}\br_{kk}=\hr_{k-1,k-1}\hr_{kk}$ and $\hr_{k-1,k-1}<\br_{k-1,k-1}$ (see the proof of Lemma \ref{l:probsr}), we have  $\br_{kk}<\hr_{kk}$, completing the proof. \quad $\Box$
\medskip

From Lemmas \ref{t:compper} and \ref{l:compsr} we immediately obtain the following result.

\begin{theorem}\label{t:complll}
Suppose that the ILS problem \eqref{e:ILSR} is transformed to the ILS problem \eqref{e:reduced},
where $\bbR$ is obtained by Algorithm \ref{a:LLL}.
Then
\beqnn
\hat{\zeta}(\R) \geq \hat{\zeta}(\bbR),
\eeqnn
where the equality holds if and only if no column permutation occurs during the LLL reduction process.
Any size reductions on the superdiagonal entries of $\R$ which is immediately followed by a column permutation
during the LLL reduction process will reduce the   complexity $\hat{\zeta}$.
All other size reductions have no effect on  $\hat{\zeta}$.
 \end{theorem}
\medskip

The result on the effect of the size reductions is consistent with a result given in  \cite{XieCA11},
which shows that all the size reductions on the off-diagonal entries
above the superdiagonal of $\R$ and the  size reductions on the  superdiagonal entries of $\R$
which are not followed by column permutations
have no effect on  the search speed of the Schnorr-Euchner algorithm for finding the ILS solution.

Like Theorem \ref{t:2delta} in Section \ref{s:delta} we can show that when $n=2$
larger $\delta$ will decrease the complexity $\hat{\zeta}$ more,
but when $n\geq 3$,  it may not be true, although our simulation results indicated that usually it is true.

In Section \ref{s:ub} we gave some upper bounds on the success probability
of the Babai point after the LLL reduction.
Here we can use \eqref{in:lub} to give a lower bound on the complexity $\hat{\zeta}$ after the LLL reduction.
To save space, we will not give any details.

\section{Summary and future work} \label{s:sum}

We  have shown that the success probability $P_\sB$ of the Babai point will increase
and the complexity $\hat{\zeta}$ of sphere decoders will decrease if the LLL reduction algorithm
given in Algorithm \ref{a:LLL} is applied for lattice reduction.
We have also discussed how the parameter $\delta$ in the LLL reduction affects  $P_\sB$ and $\hat{\zeta}$.
Some upper bounds on $P_\sB$ after the LLL reduction have been presented.
In addition,  we have shown that $P_\sB$ is a better lower bound on the success probability of ILS estimator
than the lower bound given in \cite{HasB98}.

The implementation of LLL reduction is not unique.
The KZ reduction \cite{KorZ73} is also an LLL reduction.
But the KZ conditions  are stronger than the LLL conditions.
Whether some implementations of the KZ reduction can always  increase $P_\sB$ and decrease $\hat{\zeta}$
and whether the improvement is more significant compared with the regular LLL reduction algorithm
given in Algorithm \ref{a:LLL} will be studied in the future.

In this paper, we assumed the model matrix $\A$ is deterministic.
If $\A$ is a random matrix following some distribution,  what is the formula of $P_\sB$?
what is the expected value of the search complexity? and how does the LLL reduction affect them?
These questions are for future studies.

\section*{Acknowledgment}
We are  grateful to Robert Fischer and the referees for their valuable and thoughtful  suggestions.
We would also like to thank Damien Stehl\'{e} for helpful discussions and for providing a reference.

\bibliographystyle{IEEEtran}.
%\bibliography{ref-SRB}

%\begin{biography}[{\includegraphics[width=1in,height=1.25in,clip,keepaspectratio]{mshell}}]{Michael Shell}

%\begin{biography}[{\includegraphics[width=1in,height=1.25in,clip,keepaspectratio]{chang}}]{Xiao-Wen Chang}
%\end{biography}

\begin{IEEEbiographynophoto}{Xiao-Wen Chang }
is an Associate Professor in the School of Computer Science at McGill
University. He obtained his B.Sc.\ and M.Sc.\ in Computational Mathematics from Nanjing University (1986,1989)
and his Ph.D.\ in Computer Science from McGill University (1997). His research interests are in the area of scientific
computing, with particular emphasis on numerical linear algebra and its
applications. Currently he is mainly interested in parameter estimation methods,
including integer least squares, and as well as their applications in communications, signal processing
and satellite-based positioning and wireless localization.  
He has published about fifty papers in refereed journals.
\end{IEEEbiographynophoto}

\begin{IEEEbiographynophoto}{Jinming Wen}
received his Bachelor degree in Information and Computing Science from Jilin Institute of
Chemical Technology, Jilin, China, in 2008 and his M.Sc. degree in Pure Mathematics from
the Mathematics Institute of Jilin University, Jilin, China, in 2010.
He is currently pursuing a Ph.D. in The Department of Mathematics and Statistics, McGill University, Montreal. 
His research interests are in the area of integer least squares problems and 
their applications in communications and signal processing.
\end{IEEEbiographynophoto}

\begin{IEEEbiographynophoto}{Xiaohu Xie}
received his Bachelor degree in Computer Science and Technology from Wuhan University of Technology, Wuhan, China, in 2007 and his M.Sc. degree in Computer Science and Technology from Wuhan University of Technology, Wuhan, China, in 2009.
He is currently pursuing a Ph.D. in The School of Computer Science, McGill University,
Montreal. Currently  his research focuses on the theories and algorithms for  integer least squares problems.
\end{IEEEbiographynophoto}

%\newpage



%\vfill


\end{document}